\documentclass[a4paper,12pt]{article}
\usepackage{amsmath}
\usepackage{amssymb}
\title{Drinfel'd Twisted Superconformal Algebra and Structure of Unbroken Symmetries}
\author{%
Manabu Irisawa%
  \pseudolabel{a}%
  \footnote{E-mail : irisawa@kiso.phys.metro-u.ac.jp}, \
Yoshishige Kobayashi%
  \pseudolabel{b}%
  \footnote{E-mail : yosh@th.phys.titech.ac.jp}~ and \
Shin Sasaki%
  \pseudolabel{c}%
  \footnote{E-mail : shin-s@yukawa.kyoto-u.ac.jp}%
}
\date{August, 2007}
\makeatletter
\def\pseudolabel#1{\( {}^{\mathrm{#1}} \)}
\def\metadata#1{\begin{flushright} \@date \\ hep-th/#1 \end{flushright}}
\def\thetitle{\begin{center} \LARGE \textbf{\@title} \end{center}}
\def\theauthor{\begin{center} \large \@author \end{center}}
\newenvironment{address}[1]{\begin{center}\pseudolabel{#1}\!}{\end{center}}
\def\@JLone<#1,#2>{#1}
\def\@JLtwo<#1,#2,#3>{#2}
\def\@JLyear<#1,#2,#3,#4>{#3}
\def\@JLpage<#1,#2,#3,#4>{#4}
\newcommand\JL[1]{\@JLone<#1>\ {\bfseries \@JLtwo<#1>} (\@JLyear<#1>), \@JLpage<#1>}
\def\@Jpage<#1,#2,#3>{#3}
\newcommand\andvol[1]{{\bfseries \@JLone<#1>} (\@JLtwo<#1>), \@Jpage<#1>}

\newcommand\PTPS[1]{\textit{Prog.~Theor.~Phys.~Suppl.}\ {No. \@JLone<#1> (\@JLtwo<#1>), \@Jpage<#1>}}

\newcommand\PRD[1]{ \textit{Phys.~Rev.}\ {\bfseries D}\andvol{#1}}

\newcommand\PLB[1]{ \textit{Phys.~Lett.}\ {\bfseries B}\andvol{#1}}

\newcommand\IJMP[1]{\textit{Int.~J.~Mod.~Phys.}\ \andvol{#1}}

\newcommand\JP[1]{  \textit{J.~of Phys.}\ \andvol{#1}}

\newcommand\PRP[1]{ \textit{Phys.~Rep.}\ \andvol{#1}}
\newcommand\JHEP[1]{\textit{J.~High Energy Phys.}\ \andvol{#1}}
\makeatother
\begin{document}
%
\begin{titlepage}
\thispagestyle{empty}
\renewcommand{\thefootnote}{\fnsymbol{footnote}}
\vspace*{0.0cm}
\metadata{0606207}
\vspace{0.8cm}
\thetitle
\vspace{1.5cm}
\theauthor
%
\begin{address}{a}
\textit{%
Department of Physics, 
Graduate School of Science and Engineering, \\ 
Tokyo Metropolitan University, \\
Tokyo 192--0397, JAPAN
}
\end{address}
%
\begin{address}{b}
\textit{%
Department of Physics, Faculty of Science, \\
Tokyo Institute of Technology,\\
Tokyo 152--8551, JAPAN
}
\end{address}
%
\begin{address}{c}
\textit{%
Yukawa Institute for Theoretical Physics,\\
Kyoto University,\\
Kyoto 606--8502, Japan
}
\end{address}
\vspace{0.5cm}
\begin{abstract}
We investigate deformed superconformal symmetries on 
non(anti)commutative (super)spaces 
from the point of view of the Drinfel'd twisted symmetries.
We classify all possible twist elements
derived from an abelian subsector of the superconformal
algebra.
The symmetry breaking caused by the non(anti)commutativity of the (super)spaces
is naturally interpreted as the modification of
their coproduct emerging from the corresponding
twist element.
The remaining unbroken symmetries are determined
by the commutative properties of those
symmetry generators possessing the twist element.
We also comment on non-canonically deformed non(anti)commutative superspaces,
particularly those derived from the superconformal twist 
element \( \mathcal{F}_{\mathrm{SS}} \).%
\end{abstract}
\end{titlepage}
\setcounter{footnote}{0} 
%
%
\section{Introduction}
\label{SEC:intro}
The study of noncommutative spaces has recently attracted considerable
interests, it is thought that they may provide a fundamental basis for a
theory of quantum gravity \cite{NC_Field_Theory}.
 Superstring theory, which is believed to be the most promising
possibility as a consistent theory of quantum gravity, provides a realization
of noncommutative space \cite{Seiberg-Witten}. 
The simplest noncommutative space, a noncommutative plane, possesses a
so-called canonical structure among its coordinates expressed as
\begin{equation}
	[ x^{m} , x^{n} ] = i \theta^{mn} \not= 0, \label{EQN:SpacetimeNoncommutativity}
\end{equation}
where \( \theta^{mn} \) is a constant noncommutativity parameter.
Note that this 
canonical noncommutativity breaks the Lorentz invariance of the theory.
Field theories on such a noncommutative
plane have been intensively studied. (See for example
\cite{NC_Field_Theory} and references therein.)
\par
The space-time noncommutative plane has been generalized to
superspaces \cite{Boer_Grassi_Nieuwenhuizen,Seiberg}.
The supersymmetric counterpart of 
(\ref{EQN:SpacetimeNoncommutativity}) is given by
\begin{eqnarray}
\label{EQN:SuperspaceNonanticommutativity}
	\{ \theta^{\alpha}, \theta^{\beta} \} = C^{\alpha \beta} \not= 0,
\end{eqnarray}
where $\theta^{\alpha}$ is the fermionic coordinate of the superspace $(x^m, 
\theta^{\alpha}, \bar{\theta}^{\dot{\alpha}})$. In this case, the fermionic coordinate 
is not a Grassmann variable, but, instead, it satisfies a Clifford-like algebra.
In general, the non(anti)commutativity of (super)spaces breaks
a part of the symmetry of the theory.
It is known that the non(anti)commutativity of the superspace 
(\ref{EQN:SuperspaceNonanticommutativity}) breaks half of the 
supersymmetry corresponding to the anti-supercharge $\overline{Q}$
and the Lorentz symmetry $\overline{M}_{\dot{\alpha}\dot{\beta}}$ \cite{Seiberg}.
When the theory possesses superconformal symmetry,
the deformation (\ref{EQN:SuperspaceNonanticommutativity}) also breaks
the dilatation symmetry, R-symmetry and half of the superconformal
supersymmetry $\overline{S}$ \cite{Grassi_Ricci_RoblesLlana,SJR}.
However, in \cite{SJR}, it is 
pointed out that a linear combination of the dilatation and R-symmetry is 
preserved.
\par
In \cite{CKNT}, it is shown that the canonical noncommutative plane
can be described by a representation space of the Drinfel'd twisted Poincar\'e
algebra. This suggests 
a deep relation between quantized (Hopf) algebra and noncommutative geometry. 
The theory defined on the canonical noncommutative space preserves the
{\it Drinfel'd twisted Poincar\'e symmetry},
even though the ordinary Lorentz symmetry is broken.
This idea has been extended to supersymmetry and/or conformal symmetry
\cite{KS, Ihl_Saemann, Zupnik,Zupnik2, Matlock, Banerjee_Lee_Siwach}
and there are many applications
to field theories,\cite{Recent_NC_Field_Theory}
especially a noncommutative theory of gravity
\cite{Aschieri_Dimitrijevic_Meyer_Wess}.
\par
In a related work, the authors of \cite{Lukierski_Woronowicz},
proposed alternative structures of twist elements
satisfying the twist equation and showed that the proper choice 
of the twist element leads to a non-canonical 
noncommutative space, e.g. Lie algebra and a quadratic type of noncommutative 
space.
\par
In this paper, we study various aspects of broken 
symmetries on non(anti)com\-mutative (super)spaces in the sense of the 
Drinfel'd twisted Hopf algebra. Because the symmetry breaking is
caused by the twist element, we can systematically classify the 
broken and unbroken symmetries
in the non(anti)commutative (super)spaces
in the language of the twisted Hopf algebra.
As an example, we clarify the broken and unbroken symmetries on the 
superspace caused by twist elements that are constructed from the generators of
the superconformal algebra.
\par
The organization of this paper is as follows. In \S \ref{SEC:twist},
the concept of the Drinfel'd twist in the context of Hopf algebra is
introduced. In \S \ref{SEC:symmetries}, we classify broken and 
unbroken symmetries on the various non(anti)commutative superspaces.
In particular, we focus on the non-canonical type of 
non(anti)commutative superspaces caused by the superconformal twist 
elements and classify broken and unbroken symmetries. Section 
\ref{SEC:twisted_coproduct} is devoted to the calculation of deformed 
coproducts, and \S \ref{SEC:summary} contains a summary of this paper.
%
%
\section{The Drinfel'd twisted Hopf algebra and noncommutative spaces}
\label{SEC:twist}
In this section, we review the twisted Hopf algebra and
its representation space.
A Hopf algebra
\(
	\left( \,
		\mathcal{H}, +, \circ, \iota,
		\Delta, \epsilon, \gamma \,;
		\mathbb{K}
	\, \right)
\)
over a field \( \mathbb{K} \) is a unital associative algebra
\(
	\left( \,
		\mathcal{H}, + , \circ, \iota \,;
		\mathbb{K}
	\, \right)
\)
with the following linear maps;
\begin{equation}
\begin{array}{rccl}
	\text{addition} &
	+ & : &
	\mathcal{H} \times \mathcal{H} \to \mathcal{H}
\cr
	\text{product} &
	\circ &	: &
	\mathcal{H} \times \mathcal{H} \to \mathcal{H}
\cr
	\text{unit} &
	\iota & : &
	\mathbb{K} \to \mathcal{H}
\end{array}
\qquad
\begin{array}{rccl}
	\text{coproduct} &
	\Delta & : &
	\mathcal{H} \to \mathcal{H} \otimes \mathcal{H}
\cr
	\text{counit} &
	\epsilon & : &
	\mathcal{H} \to \mathbb{K}
\cr
	\text{antipode} &
	\gamma & : &
	\mathcal{H} \to \mathcal{H}
\end{array}.
\end{equation}
These maps possesses some properties of duality, as we now describe.
Using the Sweedler notation,
\(
	\Delta(a)
	=	\sum_{i} a_{1}^{i} \otimes a_{2}^{i} = \bar{a}_{i} \otimes \bar{a}^{i},
\)
for the coproduct, the relations between the algebra and coalgebra
\begin{alignat}{7}
\label{EQN:HopfAlgebraStructures}
	\text{coassociativity}
	& : \ & &
	\Delta( \bar{a}_{i} ) \otimes \bar{a}^{i}
	=	\bar{a}_{i} \otimes \Delta( \bar{a}^{i} )
& &
	\longleftrightarrow \
	\left( a \circ b \right) \circ c
	=	a \circ \left( b \circ c \right)
	& & \, : & & \
	\text{associativity}
\nonumber \\
	\text{counit}
	& : \ & & \,
	\epsilon( \bar{a}_{i} ) \, \bar{a}^{i}
	=	\bar{a}_{i} \, \epsilon( \bar{a}^{i} )
	=	a
& &
	\longleftrightarrow \
	\iota( k ) \circ a
	=	a \circ \iota( k )
	=	k a
	& & \, : & & \
	\text{unit}
\\
	\text{antipode}
	& : \ & & \,
	\gamma( \bar{a}_{i} ) \circ \bar{a}^{i}
	=	\bar{a}_{i} \circ \gamma( \bar{a}^{i} )
	=	\iota( \epsilon( a ) )
\hspace{-3em}
\nonumber
\end{alignat}
hold for all \( a,b,c \in \mathcal{H} \), \( k \in \mathbb{K} \).
We denote the unit element of the Hopf algebra
by \( {\bf 1}  \).
The product $\circ$ is extended to tensor powers,
\(
	\circ
	:	\mathcal{H}^{\otimes n} \! \times \mathcal{H}^{\otimes n}
		\to	\mathcal{H}^{\otimes n}
\),
which is written in concrete form as
\begin{equation}
\label{EQN:ExpandedProduct}
	\left( \, a \otimes \cdots \otimes b \, \right)
	\circ
	\left( \, c \otimes \cdots \otimes d \, \right)
	:=	\left( \, a \circ c \, \right)
		\otimes \cdots \otimes
		\left( \, b \circ d \, \right)
\quad
	\forall a, b, c, d \in \mathcal{H}.
\end{equation}
(Hereafter, the symbol \( \circ \) will be omitted unless otherwise noted.)
We require the maps to satisfy the following relations:
\begin{equation}
\label{EQN:CompatibilityForAllMaps}
	\Delta( a b )
	=	\Delta( a ) \Delta( b )
\, , \quad
	\epsilon( a b )
	=	\epsilon( a ) \, \epsilon( b )
\, , \quad
	\gamma( a b )
	=	\gamma( b ) \gamma( a ).
\end{equation}
As can be seen from the last relations in eqs.
(\ref{EQN:HopfAlgebraStructures}) and (\ref{EQN:CompatibilityForAllMaps}),
the antipode behaves like an inverse element.
\par
A pair \( ( \rho, \mathcal{A} ) \), consisting of a representation \( 
\rho \) and a space \( \mathcal{A} \)
determine how \( \mathcal{H} \) acts on \( \mathcal{A} \), 
which is called a (left) \( \mathcal{H} \)-module algebra.
We define the action \( \triangleright \) by 
\begin{gather}
\label{EQN:Action}
	\left( \, a \otimes \cdots \otimes b \, \right)
	\triangleright
	\left( \, \phi \otimes \cdots \otimes \psi \, \right)
	:=	\rho( a ) \, \phi
		\otimes \cdots \otimes
		\rho( b ) \, \psi \, ,
\\
\label{EQN:Coaction}
	( a b )
	\triangleright \phi
	=	\rho( a b ) \, \phi
	=	\rho( a ) \, \rho( b ) \, \phi
	=	a \triangleright
		\left( \, b \triangleright \phi \, \right),
\end{gather}
for all \( a, b \in \mathcal{H} \) and \( \phi, \psi \in \mathcal{A} \).
In addition, we require an algebraic structure on \( \mathcal{A} \),
i.e.~the multiplication denoted by
\(
	\mu \left( \, \phi \otimes \psi \, \right)
	=:	\phi \cdot \psi,
\)
which is compatible with a coproduct on the Hopf algebra:
\begin{equation}
\label{EQN:ModuleAlgebra}
	a \triangleright
	\mu \left( \, \phi \otimes \psi \, \right)
	=	a \triangleright \left( \, \phi \cdot \psi \, \right)
	=	\left( \, \bar{a}_i \triangleright \phi \, \right)
		\cdot
		\left( \, \bar{a}^i \triangleright \psi \, \right)
	=	\mu \left( \,
			\Delta( a ) \triangleright
			\left( \, \phi \otimes \psi \, \right)
		\, \right).
\end{equation}
For more details, see \cite{Hopf_Algebra}.
\par
We can systematically construct a modified Hopf algebra
\(
	\left( \,
		\mathcal{H}, +, \circ, \iota,
		\Delta_{t}, \epsilon, \gamma_{t} \,;
		\mathbb{K}
	\, \right)
\)
from an arbitrary, given Hopf algebra
\(
	\left( \,
		\mathcal{H}, +, \circ, \iota,
		\Delta, \epsilon, \gamma \,;
		\mathbb{K}
	\, \right)
\)
\cite{Drinfeld} as we now demonstrate.
First, assume that there exists an invertible element
\(
	\mathcal{F}
	=	\sum_{i} f_{\mathrm{L}}^{i} \otimes f_{\mathrm{R}}^{i}
		\in \mathcal{H} \otimes \mathcal{H}
\),
called a {\it twist element},
which satisfies the following relations:
\begin{align}
\label{EQN:TwistEquation}
&	\mathcal{F}_{12}
	\left( \, \Delta \otimes \mathrm{id} \, \right) \mathcal{F}
	=	\mathcal{F}_{23}
		\left( \, \mathrm{id} \otimes \Delta \, \right) \mathcal{F}
\quad
	\bigl( \,
		\mathcal{F}_{12}
		:=	\mathcal{F} \otimes {\bf 1}
	\, , \
		\mathcal{F}_{23}
		:=	{\bf 1} \otimes \mathcal{F}
	\, \bigr),
\\
\label{EQN:CounitCondition}
&	\left( \, \epsilon \otimes \mathrm{id} \, \right) \mathcal{F}
	=	\left( \, \mathrm{id} \otimes \epsilon \, \right) \mathcal{F}
	=	{\bf 1}.
\end{align}
Then the new maps of the modified algebra are given by
\begin{alignat}{2}
\label{EQN:TwistedCoproduct}
	\Delta_{t}
&	:=	\mathrm{Ad}_{\mathcal{F}} \, \Delta
\, , \quad
&	\Delta_{t}( a )
&	=	\mathcal{F} \, \Delta( a ) \, \mathcal{F}^{-1},
\\
\label{EQN:TwistedAntipode}
	\gamma_{t}
&	:=	\mathrm{Ad}_{U} \, \gamma
\, , \quad \
&	\gamma_{t}(a)
&	=	U \gamma( a ) \, U^{-1}
\qquad
	\forall a \in \mathcal{H}.
\end{alignat}
Here, \( \mathcal{F}^{-1} \in \mathcal{H} \otimes \mathcal{H} \) is the inverse
of \( \mathcal{F} \) satisfying
\(
	\mathcal{F} \mathcal{F}^{-1}
	=	\mathcal{F}^{-1} \mathcal{F}
	=	{\bf 1} \otimes {\bf 1}
\)
and \( U := {\textstyle \sum_{i}} \, f_{\mathrm{L}}^{i} \, \gamma( f_{\mathrm{R}}^{i} ) \).
All other maps unchanged.
This Hopf algebra is referred to as the {\it Drinfel'd twisted} or 
merely the {\it twisted} Hopf algebra.
A representation space of the twisted Hopf algebra
\( \mathcal{A}_{t} \)
is derived from the structure of a given \( \mathcal{H} \)-module \( \mathcal{A} \),
along with the manner in which the twisted \( \mathcal{H} \) is constructed from the untwisted \( \mathcal{H} \).
The representation \( \rho \) of \( \mathcal{H} \) is the same, because
the twist element does not affect the algebraic maps \( \left( \, +, \circ \, \right) \).
The only difference regards the multiplication on the module algebra,
\begin{equation}
\label{EQN:TwistedMultiplication}
	\phi \star \psi
	=	\mu_{t} \left( \, \phi \otimes \psi \, \right)
	:=	\mu
		\left( \,
			\mathcal{F}^{-1} \triangleright
			\left( \, \phi \otimes \psi \, \right)
		\, \right).
\end{equation}
This {\it twisted multiplication} \( \mu_{t} \)
is compatible with the twisted coproduct \( \Delta_{t} \),
\begin{align}
\label{EQN:TwistedModuleAlgebra}
	a \triangleright
	\mu_{t} \left( \, \phi \otimes \psi \, \right)
&	=	\mu
		\left( \,
			\Delta( a ) \triangleright
			\mathcal{F}^{-1} \triangleright
			\left( \, \phi \otimes \psi \, \right)
		\, \right)
\nonumber \\
&	=	\mu
		\left( \,
			\mathcal{F}^{-1} \triangleright
			\Delta_{t}( a ) \triangleright
			\left( \, \phi \otimes \psi \, \right)
		\, \right)
\nonumber \\
&	=	\mu_{t}
		\left( \,
			\Delta_{t}( a ) \triangleright
			\left( \, \phi \otimes \psi \, \right)
		\, \right).
\end{align}
\par
A universal enveloping algebra is an example of a Hopf algebra.
The universal enveloping Poincar\'e algebra is frequently considered 
in connection with the noncommutative plane \cite{CKNT, KS, Zupnik2}.
Generally speaking, Lie algebras are neither unital nor associative.
The structure of the Hopf algebra requires both.
Therefore we should regards a Lie algebra \( \mathfrak{g} \) as a vector space
and introduce the formal product on it with the unit element defined 
as the zeroth power of the product, as follows:
\begin{equation}
	u_1 \! \circ \cdots \circ u_n
	:= u_1 \! \cdots u_n \in \mathfrak{g}^n
\
	\left( \, u_i \in \mathfrak{g} \, , \; n \ge 2 \, \right)
\, , \quad
	{\bf 1} \in \mathfrak{g}^0 \simeq \mathbb{C}.
\end{equation}
The direct sum of all ranks of the tensor products,
\( \mathcal{T}(\mathfrak{g}) = \bigoplus_{n=0}^{\infty} \mathfrak{g}^n \),
is called a tensor algebra.
The tensor algebra \( \mathcal{T}(\mathfrak{g}) \) has no Lie algebraic structure.
To define the Lie algebraic property, we impose the condition that
the commutator can be replaced by the Lie bracket,
\begin{equation}
	u \cdots v \, ( \, w \, x - x \, w \, ) \, y \cdots z
	\sim
	u \cdots v \, [ \, w , x \, ] \, y \cdots z
\quad
	\forall u,v,w,x,y,z \in \mathfrak{g}.
\end{equation}
Mathematically, the ideal \( \mathcal{I} \) generated by
\( u \, v - v \, u - [ \, u , v \, ] \) for all \( u , v \in \mathfrak{g} \)
gives the universal enveloping algebra as a quotient space,
\( \mathcal{U}(\mathfrak{g}) = \mathcal{T}(\mathfrak{g}) / \mathcal{I} \).
 
Let us consider a concrete example.
The Poincar\'e algebra \( \mathcal{P} \) consists of the translation
generators $P_{m}$ and the Lorentz generators $M_{mn}$, 
with the following commutation relations:
\begin{align}
\label{EQN:PoincareAlgebra}
	\left[ \, P_{m} , P_{n} \, \right]
&	=	0 \, ,
\nonumber \\
	\left[ \, M_{mn} , M_{kl} \, \right]
&	=
	-	i \eta_{mk} M_{nl}
	+	i \eta_{nk} M_{ml}
	+	i \eta_{ml} M_{nk}
	-	i \eta_{nl} M_{mk} \, ,
\nonumber \\
	\left[ \, M_{mn} , P_{k} \, \right]
&	=	-i \eta_{mk} P_{n} + i \eta_{nk} P_{m} \, .
\end{align}
Here, \( \eta_{mn} = \mathrm{diag}(-,+,+,+) \) represents the Minkowski metric.
The Poincar\'e algebra \( \mathcal{P} \) becomes
a Hopf algebra through the definitions
\begin{alignat}{7}
\label{EQN:HopfAlgebra_Poincare}
	&	\mathcal{H} = \mathcal{U}(\mathcal{P})
	\, , \quad
		\mathbb{K} = \mathbb{C} \, ,
	& & & & & &
	\nonumber \\
	&	\Delta(u) := u \otimes {\bf 1} + {\bf 1} \otimes u \, ,
	& \quad &
		\epsilon(u) := 0 \, ,
	& \quad &
		\gamma(u) := - u
	& \quad &
		\quad \forall u \in \mathcal{P} \, ,
	\nonumber \\
	&	\Delta({\bf 1}) := {\bf 1} \otimes {\bf 1} \, ,
	& \quad &
		\epsilon({\bf 1}) := {\bf 1} \, ,
	& \quad &
		\gamma({\bf 1}) := {\bf 1} \, ,
	& \quad &
		\quad {\bf 1} \in \mathcal{U}(\mathcal{P}) \, .
\end{alignat}
Now, we consider the usual differential representation,
\( \rho : \mathcal{U}(\mathcal{P}) \to \mathrm{diff}(\mathcal{A}) \).
The \( \mathcal{U}(\mathcal{P}) \)-module algebra is the set
\( \mathcal{A} = C^{\infty}(\mathbb{R}^{1,3}) \), i.e. all smooth functions on Minkowski space.
In this case, the compatibility condition (\ref{EQN:ModuleAlgebra}) of the action
 simply constitute the Leibniz rule.
A twisted Poincar\'e algebra \( \mathcal{U}_t(\mathcal{P}) \) is 
constructed from a twist element with {\it abelian generators} $P_m$ in the algebra
\begin{equation}
\label{EQN:PP_TwistElement}
	\mathcal{F}_{\mathrm{PP}}
	:=	\exp \left(\, \frac{i}{2} \theta^{mn} P_{m} \otimes P_{n} \,\right)
	\; \in \;
	\mathcal{U}(\mathcal{P}) \! \otimes \mathcal{U}(\mathcal{P}),
\end{equation}
where \( \theta^{mn} \in \mathbb{R} \) is a constant with antisymmetric
indices%
\footnote{
	Actually, it is not necessary for \( \theta^{mn} \) to be real or even 
	with antisymmetric indices at the algebraic level.
	Even in the case that $\theta^{mn}$ is not real or antisymmetric,
	the twist element (\ref{EQN:PP_TwistElement}) 
	satisfies the twist equation (\ref{EQN:TwistEquation}).
	The reality condition is needed only after requiring 
	the consistency of the commutation relation on the representation space.
}
and \( P_m \) are the translation generators of the Poincar\'e algebra.
The element (\ref{EQN:PP_TwistElement}) satisfies the twist equation (\ref{EQN:TwistEquation})
and the counit condition (\ref{EQN:CounitCondition}) by virtue of the abelian
property of \( P_{m}\).
The explicit form of the twisted coproduct for translations and Lorentz transformations is
\begin{alignat}{2}
&	\Delta_{t}^{\mathrm{PP}}(P_{m})
&&	=	\Delta(P_{m}) \, ,
\\
&	\Delta_{t}^{\mathrm{PP}}(M_{mn})
&&	=	\Delta(M_{mn})
	-	\theta^{kl}
		\left[ \,
			\eta_{k(m} P_{n)} \otimes P_{l}
		+	P_{k} \otimes \eta_{l(m} P_{n)}
		\, \right] \, ,
\end{alignat}
where the brackets $()$ represent symmetrization of the indices.
The antipode is not modified; i.e. \( \gamma_{t}^{\mathrm{PP}} = \gamma \).
 
For space-time coordinates \( \{ x^{m} \} \) and the representation
\( \rho(P_{m}) = i \partial_{m} \, \),
we have
\begin{align}
	x^{m} \star \, x^{n}
&	:=	\mu
		\left( \,
			{ \mathcal{F}_{\mathrm{PP}} }^{-1}
			\triangleright
			\left( \,
				x^{m} \! \otimes x^{n}
			\, \right)
		\, \right)
\nonumber \\
& \	=	\mu
		\left( \,
			\exp
			\left( \,
			-	\frac{i}{2} \theta^{kl}
				\rho( P_{k} ) \otimes \rho( P_{l} ) \, \right)
				\left( \, x^{m} \! \otimes x^{n} \, \right)
			\, \right)
\nonumber \\
& \	=	\mu
		\left( \,
			\exp
			\left( \,
			\frac{i}{2} \theta^{kl} \,
				\partial_{k} \otimes \partial_{l}
			\, \right)
			\left( \, x^{m} \! \otimes x^{n} \, \right)
		\, \right)
	=	x^{m} x^{n}
	+	\frac{i}{2} \theta^{mn}.
\end{align}
We have thus seen that the noncommutative plane
\(
	[ \, x^{m} , x^{n} \, ]_{\star}
	=	\frac{i}{2} \theta^{mn} - \frac{i}{2} \theta^{nm}
	=	i \theta^{mn}
\)
is realized by the twist element (\ref{EQN:PP_TwistElement}).
\par
We can consider a Hopf superalgebra
in almost the same manner as the Poincar\'e algebra above. There exist twist elements
which induce a Moyal-type multiplication on the superspace \cite{KS}.
In fact,
the \( \mathcal{N}=1 \) super Poincar\'e algebra \( \mathcal{SP} \)
has the structure of a Hopf algebra with a graded version of the product:
\begin{equation}
	\left( \, a \otimes b \, \right) \left( \, c \otimes d \, \right)
	=	(-1)^{|b||c|} \, a \, c \otimes b \, d
\qquad
	\forall \, a, b, c, d \in \mathcal{SP}.
\end{equation}
Here, $|b|$ represents the fermionic character of $b$, which is defined as
\begin{equation}
|a| = \left\{
\begin{array}{ll}
0 & \text{if} \ a \ \text{is bosonic}, \\
1 & \text{if} \ a \  \text{is fermionic}.
\end{array}
\right.
\end{equation}
All ranks of the graded product are defined in the manner of 
(\ref{EQN:ExpandedProduct}), except for the overall sign,
which is determined by how many exchanges of fermionic elements
are involved.
When \( \mathbb{K} \) contains not only a c-number but also an anticommutative
number, i.e. a Grassmann number,  \( \mathbb{K} \) is no longer a field 
but a Grassmann number ring.
In that case, for the consistency of \( \mathbb{K} \) and
the Hopf superalgebra, we further require the following condition
for all \( k \in \mathbb{K} \) and \( a_i \in \mathcal{SP} \):
\begin{align}
	k \, a_1 \!\otimes a_2 \!\otimes a_3 \cdots
&	=	(-1)^{ |k| |a_1\!| } \,
		a_1 k \otimes a_2 \!\otimes a_3 \cdots
\nonumber \\
&	=	(-1)^{ |k| |a_1\!| } \,
		a_1 \!\otimes k \, a_2 \!\otimes a_3 \cdots
\nonumber \\
&	=	(-1)^{ |k| \left( |a_1\!| + |a_2\!| \right) } \,
		a_1 \!\otimes a_2 \!\otimes k \, a_3 \cdots.
\end{align}
The definitions (\ref{EQN:HopfAlgebra_Poincare})
are applied to all elements \( u \in \mathcal{SP} \), as well as
\( \mathcal{P} \).
The twist element constructed from the fermionic generators
\( Q \) in the super Poincar\'e algebra,
\begin{equation}
\begin{split}
	\{ \, Q_{\alpha}, Q_{\beta} \, \}
&	=	0
, \quad
	\{ \, Q_{\alpha}, \bar{Q}_{\dot{\alpha}} \, \}
	=	2 \, { \sigma^{m} }_{\alpha\dot{\alpha}} P_{m},
\\
	[ \, Q_{\alpha}, P_{m} \, ]
&	=	0
, \quad
	[ \, Q_{\alpha}, M_{mn} \, ]
	=	- i \, { ( \sigma_{mn} )_{\alpha} }^{\beta} Q_{\beta},
\end{split}
\quad
\text{and the h.c. counterparts}
\ ,
\end{equation}
with a constant symmetric matrix \( C^{\alpha\beta} \in \mathbb{C} \),
\begin{equation}
\label{EQN:QQ_TwistElement}
	\mathcal{F}_{\mathrm{QQ}}
	:=	\exp
		\left( \,
			- \frac{1}{2} C^{\alpha\beta}
			Q_{\alpha} \!\otimes Q_{\beta}
		\, \right),
\end{equation}
satisfies eqs.(\ref{EQN:TwistEquation}) and (\ref{EQN:CounitCondition}).
Here we adopt the representation
 \( \rho( Q_{\alpha} ) \, \theta^{\beta} = i { \delta_{\alpha} }^{\beta} \)
for the fermionic coordinates \( \{ \theta^{\beta} \} \).
The twisted multiplication is
\begin{align}
	\theta^{\alpha} \star\, \theta^{\beta}
&	:=	\mu
		\left( \,
			{ \mathcal{F}_{\mathrm{QQ}} }^{-1} \triangleright
			\left( \,
				\theta^{\alpha} \!\otimes \theta^{\beta}
			\, \right)
		\, \right)
\nonumber \\
& \	=	\mu
		\left( \,
			\exp
			\left( \,
				\frac{1}{2} \, C^{\gamma\delta}
				\rho( Q_{\gamma} ) \otimes \rho( Q_{\delta} )
			\, \right)
			\left( \,
				\theta^{\alpha} \!\otimes \theta^{\beta}
			\, \right)
		\, \right)
\nonumber \\
& \	=	\mu
		\left( \,
			\theta^{\alpha} \!\otimes \theta^{\beta}
		-	\frac{1}{2} \, C^{\gamma\delta}
			\left[ \, \rho( Q_{\gamma} ) \, \theta^{\alpha} \, \right]
			\otimes
			\left[ \, \rho( Q_{\delta} ) \, \theta^{\beta} \, \right]
		\, \right)
	=	\theta^{\alpha} \theta^{\beta}
	+	\frac{1}{2} \, C^{\alpha\beta}.
\end{align}
Therefore, we find the canonical non-anticommutativity of the superspace
\(
	\{ \, \theta^{\alpha}, \theta^{\beta} \, \}_{\star}
	=	C^{\alpha\beta}
\)
.
The coproducts \( \Delta_{t}^{\mathrm{QQ}} \) of the generators in \( \mathcal{SP} \)
are evaluated in \cite{KS}.
\par
Next, we consider other examples.
The special conformal generator \( K \) and the conformal supercharge
\( S \) are both possibilities as the twist element in the 
superconformal algebra \( \mathcal{SC} \).
The  \( K \)-\( K \) twist element \( \mathcal{F}_{\mathrm{KK}} \)
causes very complicated noncommutativities among the superspace coordinates.
Actually, the commutator of the coordinates results in an infinite series
expansion with respect to the noncommutativity parameter.
It is difficult to obtain the closed form of the commutator.
Therefore, here we consider a twist with the \( S \)-supercharges
\begin{align}
\label{EQN:SS_TwistElement}
&	\mathcal{F}_{\mathrm{SS}}
	:=	\exp
		\left( \,
			 \frac{1}{2} C^{\alpha\beta}
			S_{\alpha} \!\otimes S_{\beta}
		\, \right),
\quad
	C^{\alpha\beta}
	=	C^{\beta\alpha} \in \mathbb{C} \, ,
\\
\label{EQN:S_Supercharge}
&	\rho( S_{\alpha} )
	=	- i \left( \,
			{ x_{\alpha} }^{\dot{\beta}} \, \theta^{\beta} \sigma^{m}_{\beta\dot{\beta}}
		-	i \theta^2 \sigma^{m}_{\alpha\dot{\beta}} \, \bar{\theta}^{\dot{\beta}}
		\, \right)
		\partial_{m}
	+	2 \, i \, \theta^2 \partial_{\alpha}
	+	\left( \,
			{ x_{\alpha} }^{\dot{\beta}}
		+	2 \, i \, \theta_{\alpha} \bar{\theta}^{\dot{\beta}}
		\, \right)
		\bar{\partial}_{\dot{\beta}} \, .
\end{align}
This twist element induces the non-canonical commutation relation%
\footnote{%
To properly express the totally (anti)symmetrized terms 
we should employ the by star product
on the right-hand side of eq.(\ref{EQN:SS_TwistedCommutator}).
However, the additional terms included by so doing would not change the 
result presented in eq.(\ref{EQN:SS_TwistedCommutator}).
}%
\begin{align}
\label{EQN:SS_TwistedCommutator}
&	[ \, x^{m}, x^{n} \, ]_{\star}
	=	 C^{\alpha\gamma}
		\bigl(
			{ x_{\alpha} }^{\dot{\beta}} \!\star
			{ x_{\gamma} }^{\dot{\delta}} \!\star
			\theta^{\beta} \!\star \theta^{\delta}
		\bigr)
		{ \sigma^{m} }_{\beta\dot{\beta}} \, { \sigma^{n} }_{\delta\dot{\delta}} \, .
\end{align}
Unlike the case of the \( K \)-\( K \) twist,
the \( S \)-supercharge twist element induces
noncommutativities only at first order in the noncommutativity parameter.
The higher order terms vanish
because of the nilpotency of the Grassmann coordinates \( \{ \theta^{\alpha} \} \).
%
%
\section{Classification of broken and unbroken symmetries}
\label{SEC:symmetries}
In this section, we study the symmetry breaking
on non(anti)commutative superspaces and its relation to
the twisted Hopf algebra.
\par
Modification of the superconformal symmetry
on the canonical type of non-(anti)commutative
superspace (\ref{EQN:SuperspaceNonanticommutativity})
is studied in \cite{Grassi_Ricci_RoblesLlana} and \cite{SJR}.
In \cite{SJR}, it is found that the $R$-symmetry and the dilatation symmetry
are broken, but that a linear combination of these two symmetries 
remains unbroken as the new dilatation symmetry.
To show that, the authors of \cite{SJR}
calculated the commutators of the generators
with the star product explicitly.  
\par
Actually, from the point of view of the twisted algebra,
it is obvious what symmetry is broken,\footnote{
In this section, we use the terms ``broken'' and ``unbroken'' with 
regards to symmetry in the usual sense. 
Specifically, they do not refer to the twisted symmetry.
} or what combination of
 symmetries is unbroken under the non-(anti)commutative deformation of (super)space.
\par
The Drinfel'd twist modifies the multiplication rule in the representation
space, and simultaneously changes the coproduct.
This modification of the coproduct in turn changes
the action on the product of the representation, and this ensures that
the structure of the algebra is preserved.
Thus, symmetry breaking emerges as the appearance of extra terms of 
the modified coproduct, which include the noncommutativity parameters.
Inspecting there extra terms in the coproduct, 
it is clear whether any given symmetry $G$ is broken or not.
For instance, in the $Q$-$Q$ twisted case, the coproduct corresponding
to a symmetry generator $G$ becomes the following:
\begin{align}
	\Delta_{t}^{\mathrm{QQ}} (G)
&	=	G \otimes \mathbf{1} + \mathbf{1} \otimes G
\nonumber \\
&	-	\frac{1}{2} C^{\alpha \beta}
		\left( \,
			Q_{\alpha} \!\otimes [ \, Q_{\beta} , G \, ]_{\pm}
		+	(-1)^{|G|} \,
			[ \, Q_{\alpha} , G \, ]_{\pm} \otimes Q_{\beta}
		\, \right)
\nonumber \\
&	-	\frac{1}{8} C^{\gamma \delta} C^{\alpha \beta}
		\left( \,
			Q_{\gamma} Q_{\alpha} \!\otimes [ \, Q_{\delta} , [ \, Q_{\beta} , G \, ]_{\pm} \, ]_{\pm}
		+	[ \, Q_{\gamma} , [ \, Q_{\alpha} , G \, ]_{\pm} \, ]_{\pm} \otimes Q_{\delta} Q_{\beta}
		\, \right),
\label{coproductG}
\end{align}
where $[A,B]_{\pm}$ is the (anti)commutator bracket,
which gives $A B - (-1)^{|A||B|} B A$.
The terms of order \( \mathcal{O}(C^3) \) vanish because \( Q \)
is nilpotent.
From eq.(\ref{coproductG}), it is seen that the problem of determining 
whether or not the symmetry is broken or not reduces to that of 
determining whether or not the commutator $[Q,G]_{\pm}$ is nonzero.
\par
We focus on the \( \mathcal{N} = 1 \) superconformal algebra.
Possible abelian twist elements\footnote{
In fact, $M_{12}$ and $P_3$ commute,
but this combination is not considered here.
}
 in the superconformal algebra are listed in Table \ref{TBL:PossibleCombinations}.
\begin{table}[t]
\caption{Possibilities for abelian twist elements.}
\label{TBL:PossibleCombinations}
\def\enable{\( \bigcirc \)}
\renewcommand\arraystretch{1.1}
\begin{center}
\begin{tabular}{c|c|c|c|c|c|c|c|c|c}
\hline \hline
	Basic Element
	& \multicolumn{9}{|c}{Available Combinations} \cr
\hline
	\( \mathcal{F}_{\mathrm{PP}} \)
	& \enable & \enable &  &  &  &  &  & \enable &  \cr
\hline
	\( \mathcal{F}_{\mathrm{PQ}} \)
	& \enable &  &  &  &  &  &  &  &  \cr
\hline
	\( \mathcal{F}_{\mathrm{QQ}} \)
	& \enable &  & \enable &  &  &  &  &  &  \cr
\hline
	\( \mathcal{F}_{\mathrm{P\bar{Q}}} \)
	&  & \enable &  &  &  &  &  &  &  \cr
\hline
	\( \mathcal{F}_{\mathrm{\bar{Q}\bar{Q}}} \)
	&  & \enable &  & \enable &  &  &  &  &  \cr
\hline
	\( \mathcal{F}_{\mathrm{Q\bar{S}}} \)
	&  &  & \enable &  &  &  &  &  &  \cr
\hline
	\( \mathcal{F}_{\mathrm{\bar{Q}S}} \)
	&  &  &  & \enable &  &  &  &  &  \cr
\hline
	\( \mathcal{F}_{\mathrm{\bar{S}\bar{S}}} \)
	&  &  & \enable &  & \enable &  &  &  &  \cr
\hline
	\( \mathcal{F}_{\mathrm{K\bar{S}}} \)
	&  &  &  &  & \enable &  &  &  &  \cr
\hline
	\( \mathcal{F}_{\mathrm{SS}} \)
	&  &  &  & \enable &  & \enable &  &  &  \cr
\hline
	\( \mathcal{F}_{\mathrm{KS}} \)
	&  &  &  &  &  & \enable &  &  &  \cr
\hline
	\( \mathcal{F}_{\mathrm{KK}} \)
	&  &  &  &  & \enable & \enable &  &  & \enable \cr
\hline
	\( \mathcal{F}_{\mathrm{DD}} \)
	&  &  &  &  &  &  & \enable &  &  \cr
\hline
	\( \mathcal{F}_{\mathrm{DR}} \)
	&  &  &  &  &  &  & \enable &  &  \cr
\hline
	\( \mathcal{F}_{\mathrm{RR}} \)
	&  &  &  &  &  &  & \enable & \enable & \enable \cr
\hline
	\( \mathcal{F}_{\mathrm{PR}} \)
	&  &  &  &  &  &  &  & \enable &  \cr
\hline
	\( \mathcal{F}_{\mathrm{KR}} \)
	&  &  &  &  &  &  &  &  & \enable \cr
\hline
\end{tabular}
\end{center}
\end{table}
That table presents the twist elements with our notation,
\(
	\mathcal{F}_{\mathrm{XY}}
	=	\exp \,
		\bigl( \,
			C^{AB} \bigl[ \, X_{A} \otimes Y_{B} - (-1)^{|X||Y|} \, Y_{B} \otimes X_{A} \, \bigr]
		\, \bigr)
\),
and each column represents the simultaneous 
utilization of the indicated \( \mathcal{F}\), namely the possible combinations of 
abelian twist elements.
In fact, the product of the indicated elements, 
\( \mathcal{F} := \mathcal{F}_{\mathrm{PP}} \mathcal{F}_{\mathrm{PQ}} 
\mathcal{F}_{\mathrm{QQ}} \), 
satisfies the conditions (\ref{EQN:TwistEquation}) and (\ref{EQN:CounitCondition})
because the generators in their twist elements are in an abelian subsector.
\par
Let us consider the unbroken symmetries in each twist element and the corresponding 
non(anti)commutative superspaces.
Below we list the twists and the relevant commutators in the spinor 
representation.
\par
\begin{itemize}
\item {\slshape Relevant commutators in the \( P \text{-} P \) twist:}
\begin{alignat}{7}
\label{EQN:Algebra_AssociatedWith_P}
&	[ \, P_{\alpha\dot{\alpha}}, P_{\beta\dot{\beta}} \, ]
& &	=	0,
\nonumber \\
&	[ \, P_{\alpha\dot{\alpha}}, M_{\beta\gamma} \, ]
& &	=	- i \epsilon_{\alpha(\beta} P_{\gamma)\dot{\alpha}},
 \quad &
&	[ \, P_{\alpha\dot{\alpha}}, \bar{M}_{\dot{\beta}\dot{\gamma}} \, ]
& &	=	i P_{\alpha(\dot{\beta}} \epsilon_{\dot{\gamma})\dot{\alpha}},
\nonumber \\
&	[ \, P_{\alpha\dot{\alpha}}, Q_{\beta} \, ]
& &	=	0,
 \quad &
&	[ \, P_{\alpha\dot{\alpha}}, \bar{Q}_{\dot{\beta}} \, ]
& &	=	0,
\nonumber \\
&	[ \, P_{\alpha\dot{\alpha}}, S_{\beta} \, ]
& &	=	2 \, \epsilon_{\alpha\beta} \, \bar{Q}_{\dot{\alpha}},
 \quad &
&	[ \, P_{\alpha\dot{\alpha}}, \bar{S}_{\dot{\beta}} \, ]
& &	=	- 2 \, \epsilon_{\dot{\alpha}\dot{\beta}} \, Q_{\alpha},
\nonumber \\
&	[ \, P_{\alpha\dot{\alpha}}, K_{\beta\dot{\beta}} \, ]
& &	=	4i
		\left( \,
			\epsilon_{\alpha\beta} \bar{M}_{\dot{\alpha}\dot{\beta}}
		+	\epsilon_{\dot{\alpha}\dot{\beta}} M_{\alpha\beta}
		-	\epsilon_{\alpha\beta} \epsilon_{\dot{\alpha}\dot{\beta}} D
		\, \right),
\hspace{-3em}
\nonumber \\
&	[ \, P_{\alpha\dot{\alpha}}, D \, ]
& &	=	i P_{\alpha\dot{\alpha}},
 \quad &
&	[ \, P_{\alpha\dot{\alpha}}, R \, ]
& &	=	0.
\end{alignat}
\item {\slshape Relevant (anti)commutators in the \( Q \text{-} Q \) twist:}
\begin{alignat}{7}
&	[\, Q_{\alpha}, P_{\beta\dot{\beta}} \, ]
& &	=	0,
\nonumber \\
&	[ \, Q_{\alpha}, M_{\beta\gamma} \, ]
& &	=	- i \epsilon_{\alpha(\beta} Q_{\gamma)},
 \quad &
&	[ \, Q_{\alpha}, \bar{M}_{\dot{\beta}\dot{\gamma}} \, ]
& &	=	0,
\nonumber \\
& \!	\{ \, Q_{\alpha}, Q_{\beta} \, \}
& &	=	0,
 \quad &
& \!	\{ \, Q_{\alpha}, \bar{Q}_{\dot{\alpha}} \, \}
& &	=	2 P_{\alpha\dot{\alpha}},
\nonumber \\
& \!	\{ \, Q_{\alpha}, S_{\beta} \, \}
& &	=	4i M_{\alpha\beta} - 2i \epsilon_{\alpha\beta} D - 6 \epsilon_{\alpha\beta} R,
 \qquad &
& \!	\{ \, Q_{\alpha}, \bar{S}_{\dot{\beta}} \, \}
& &	=	0,
\nonumber \\
&	[\, Q_{\alpha}, K_{\beta\dot{\beta}} \, ]
& &	=	2 \, \epsilon_{\alpha\beta} \bar{S}_{\dot{\beta}},
\nonumber \\
&	[\, Q_{\alpha}, D \, ]
& &	=	\frac{i}{2} Q_{\alpha},
 \quad &
&	[\, Q_{\alpha}, R \, ]
& &	=	\frac{1}{2} Q_{\alpha}.
\end{alignat}
\item {\slshape Relevant (anti)commutators in the \( S \text{-} S \) twist:}
\begin{alignat}{7}
&	[ \, S_{\alpha}, P_{\beta\dot{\beta}} \, ]
& &	=	2 \, \epsilon_{\alpha\beta} \bar{Q}_{\dot{\beta}},
\nonumber \\
&	[ \, S_{\alpha}, M_{\beta\gamma} \, ]
& &	=	- i \epsilon_{\alpha(\beta} S_{\gamma)},
 \quad &
&	[ \, S_{\alpha}, \bar{M}_{\dot{\beta}\dot{\gamma}} \, ]
& &	=	0,
\nonumber \\
& \!	\{ \, S_{\alpha}, Q_{\beta} \, \}
& &	=	4i M_{\alpha\beta} + 2i \epsilon_{\alpha\beta} D + 6 \epsilon_{\alpha\beta} R,
 \qquad &
& \!	\{ \, S_{\alpha}, \bar{Q}_{\dot{\beta}} \, \}
& &	=	0,
\nonumber \\
& \!	\{ \, S_{\alpha}, S_{\beta} \, \}
& &	=	0,
 \quad &
& \!	\{ \, S_{\alpha}, \bar{S}_{\dot{\alpha}} \, \}
& &	=	2 K_{\alpha\dot{\alpha}},
\nonumber \\[.2em]
&	[ \, S_{\alpha}, K_{\beta\dot{\beta}} \, ]
& &	=	0,
\nonumber \\
&	[ \, S_{\alpha}, D \, ]
& &	=	- \frac{i}{2} S_{\alpha},
 \quad &
&	[ \, S_{\alpha}, R \, ]
& &	=	- \frac{1}{2} S_{\alpha}.
\end{alignat}
\item {\slshape Relevant commutators in the \( D \text{-} D \) twist:}
\begin{alignat}{14}
&	[ \, D, P_{\alpha\dot{\alpha}} \, ]
& &	=	- i P_{\alpha\dot{\alpha}},
 \quad &
&	[ \, D, M_{\alpha\beta} \, ]
& &	=	0,
 \qquad &
&	[ \, D, \bar{M}_{\dot{\alpha}\dot{\beta}} \, ]
& &	=	0,
\nonumber \hspace{5em} \\[.2em]
&	[ \, D, Q_{\alpha} \, ]
& &	=	- \frac{i}{2} Q_{\alpha},
 \quad &
&	[ \, D, \bar{Q}_{\dot{\alpha}} \, ]
& &	=	- \frac{i}{2} \bar{Q}_{\dot{\alpha}},
\nonumber \\[.2em]
&	[ \, D, S_{\alpha} \, ]
& &	=	\frac{i}{2} S_{\alpha},
 \quad &
&	[ \, D, \bar{S}_{\dot{\alpha}} \, ]
& &	=	\frac{i}{2} \bar{S}_{\dot{\alpha}},
\nonumber \\[.4em]
&	[ \, D, K_{\alpha\dot{\alpha}} \, ]
& &	=	i K_{\alpha\dot{\alpha}},
 \quad &
&	[ \, D, D \, ]
& &	=	0,
 \qquad &
&	[ \, D, R \, ]
& &	=	0.
\end{alignat}
\end{itemize}
The broken and unbroken symmetries of the twist elements
are given in Table \ref{TBL:Broken_And_Unbroken}.
\begin{table}[t]
\caption{Broken and Unbroken Symmetries}
\label{TBL:Broken_And_Unbroken}
\renewcommand\arraystretch{1.25}
\begin{center}
\begin{tabular}{|c|c||c|c|}
\hline
	Twist &
	Non(anti)commutativity &
	Broken Symmetries &
	Unbroken Symmetries \cr
\hline \hline
	\( P \)-\( P \) & 
	\( [ x^m, x^n ] = i \theta^{mn} \hfill \) &
	\( M_{\alpha\beta}, \bar{M}_{\dot{\alpha}\dot{\beta}},
	   S_{\alpha}, \bar{S}_{\dot{\alpha}},
	   K_{\alpha\dot{\alpha}}, D \) &
	\( P_{\alpha\dot{\alpha}}, Q_{\alpha}, \bar{Q}_{\dot{\alpha}}, R \) \cr
\hline
	\( Q \)-\( Q \) & 
	\( \hspace{-.1em} \{ \theta^{\alpha}, \theta^{\beta} \} = C^{\alpha\beta} \hfill \) &
	\( M_{\alpha\beta}, \bar{Q}_{\dot{\alpha}}, S_{\alpha},
	   K_{\alpha\dot{\alpha}}, D, R \) &
	\( P_{\alpha\dot{\alpha}}, \bar{M}_{\dot{\alpha}\dot{\beta}}, Q_{\alpha},
	   \bar{S}_{\dot{\alpha}}, D - iR \) \cr
\hline
	\( S \)-\( S \) & 
	\( [ x^m, x^n ] \sim C x x \theta \theta \,\footnotemark \hfill \) &
	\( P_{\alpha\dot{\alpha}}, M_{\alpha\beta}, Q_{\alpha},
	   \bar{S}_{\dot{\alpha}}, D, R \) &
	\( \bar{M}_{\dot{\alpha}\dot{\beta}},
	   \bar{Q}_{\dot{\alpha}}, S_{\alpha}, K_{\alpha\dot{\alpha}}, D - iR \) \cr
\hline
	\addtocounter{footnote}{-1}%
	\( K \)-\( K \) & 
	\( [ x^m, x^n ] \sim \Theta x x x x + \cdots \,\footnotemark \hfill \) &
	\( M_{\alpha\beta}, \bar{M}_{\dot{\alpha}\dot{\beta}},
	   Q_{\alpha}, \bar{Q}_{\dot{\alpha}},
	   P_{\alpha\dot{\alpha}}, D \) &
	\( S_{\alpha}, \bar{S}_{\dot{\alpha}}, K_{\alpha\dot{\alpha}}, R \) \cr
\hline
	\( D \)-\( D \) & 
	\( x^m \star x^n = e^{c} x^m x^n \hfill \) &
	\( P_{\alpha\dot{\alpha}},
	   Q_{\alpha}, \bar{Q}_{\dot{\alpha}}, S_{\alpha}, \bar{S}_{\dot{\alpha}},
	   K_{\alpha\dot{\alpha}} \) &
	\( M_{\alpha\beta}, \bar{M}_{\dot{\alpha}\dot{\beta}},
	   D, R \) \cr
\hline
\end{tabular}
\end{center}
\end{table}
\footnotetext{We can replace each usual product on the right-hand sides 
of the commutators by the star product.}
For the \( Q \)-\( Q \) and \( S \)-\( S \) twists,
although \( D \) and \( R \) respectively are broken individually,
the only combination \( D - iR \) survives, because 
both \( [ D , Q_{\alpha} ] - i [ R , Q_{\alpha} ] \) and 
\( [ D , S_{\alpha} ] - i [ R , S_{\alpha} ] \) are zero.
This agrees with the result obtained in \cite{SJR}.
%
%
\section{Twisted coproducts}
\label{SEC:twisted_coproduct}
The symmetry breaking on each non(anti)commutative space
can be interpreted in terms of the extra terms in the twisted coproduct.
However, in the sense of (\ref{EQN:TwistedModuleAlgebra}), 
the symmetries are not broken in terms of the twisted Hopf algebra,
i.e. modifying the multiplication and the Leibniz rule on the original (anti)commutative space.
\par
In this section, we investigate the twisted coproducts using abelian twist elements
constructed from the generators of the superconformal algebra \( \mathcal{SC} \).
The possible such twist elements are \( P \)-\( P \), \( K \)-\( K \) and \( D \)-\( D \)
for bosonic generators and \( Q \)-\( Q \) and \( S \)-\( S \) for fermionic generators.
In addition to these twist elements,
there are some mixed-type twist elements, which are composed 
of bosonic and fermionic generators.
 
Because a twist element associated with bosonic generators has an infinite
expansion series,
naively one may guess that its twisted coproduct consists
of infinitely many terms in general.
For instance, the case of eq.(\ref{EQN:PP_TwistElement}) is calculated as 
\begin{alignat}{3}
\label{EQN:PP_TwistedCoproduct}
	\Delta_{t}^{\mathrm{PP}}( G )
&	=	\Delta_{0}( G )
	+	\sum_{h=1}^{\infty} \frac{1}{h!}
		\left[ \, \prod_{j=1}^{h} \frac{i}{2} \, \theta^{m_{j}n_{j}} \right] 
\nonumber \\
&	\times	\left[ \,
			P_{m_{1}} \! \cdots P_{m_{h}}
			\!\otimes
			\mathrm{ad}_{P_{n_{1}}} \! \cdots \, \mathrm{ad}_{P_{n_{h}}} \! ( G )
		+	\mathrm{ad}_{P_{m_{1}}} \! \cdots \, \mathrm{ad}_{P_{m_{h}}} \! ( G )
			\otimes
			P_{n_{1}} \! \cdots P_{n_{h}}
		\, \right],
\end{alignat}
where \( \mathrm{ad}_{A}( B ) := [ \, A, B \, ] \) and \( G \in \mathcal{SC} \).
However, in view of eq.(\ref{EQN:Algebra_AssociatedWith_P}),
we obtain the adjoint nilpotency \( [ \, P_{m}, [ \, P_{n}, [ \, P_{k}, G \, ] \, ] \, ] = 0 \)
for all \( G \in \mathcal{SP}\).
Hence this polynomial is only second order in the parameter \( \theta \).
For the same reason, the \( K \)-\( K \) twisted coproduct has an expansion
 that terminates at second order in the noncommutativity parameter.
In fact, for the twist element 
\begin{align}
\label{EQN:KK_TwistElement}
	\mathcal{F}_{\mathrm{KK}}
	:\!&=	\exp
		\left( \,
			\Theta^{mn} K_{m} \!\otimes\! K_{n}
		\, \right),
\end{align}
we have
\begin{align}
\label{EQN:KK_TwistedCoproduct}
	\Delta_{t}^{\mathrm{KK}}( G )
&	=	\Delta_{0}( G )
	+	\Theta^{mn}
		\left( \,
			K_{m} \! \otimes [ \, K_{n} , G \, ]
		+	[ \, K_{m} , G \, ] \otimes K_{n}
		\, \right)
\nonumber \\
& \
	+	\frac{1}{2} \Theta^{kl} \Theta^{mn}
		\left( \,
			K_{k} K_{m} \! \otimes [ \, K_{l} , [ \, K_{n} , G \, ] \, ]
		+	[ \, K_{k} , [ \, K_{m} , G \, ] \, ] \otimes K_{l} K_{n}
		\, \right).
\end{align}
Here, $\Theta^{mn}$ is a non-zero noncommutativity parameter.
The \( D \)-\( D \) twist is an exceptional case.
Some coproducts have terms containing all orders of the noncommutativity parameter.
Nonetheless, by virtue of the desired property of dilatation, 
\( [ \, D, G \, ] = f_{G} \, G \), with some structure constant \( f_{G} \in \mathbb{C} \),
the \( D \)-\( D \) twisted coproduct can be written in the following simple form:
\begin{align}
\label{EQN:DD_TwistElement}
	\mathcal{F}_{\mathrm{DD}}
	:\!&=	\exp \left( \, c \, D \otimes D \, \right),
\\[-.3em]
\label{EQN:DD_TwistedCoproduct}
	\Delta_{t}^{\mathrm{DD}} ( G )
&	=	\Delta_{0} ( G )
	+	\sum_{h=1}^{\infty} \frac{c^{h}}{h!} \,
		\Bigl[ \,
			D^{h} \!\otimes \overbrace{ \mathrm{ad}_{D} \! \cdots \mathrm{ad}_{D} }^{h} ( G )
		+	\overbrace{ \mathrm{ad}_{D} \! \cdots \mathrm{ad}_{D} }^{h} ( G ) \otimes D^{h}
		\, \Bigr]
\nonumber \\
&	=	\Delta_{0} ( G )
	+	\sum_{h=1}^{\infty} \frac{(c f_{G})^{h}}{h!}
		\left( \,
			D^{h} \!\otimes G
		+	G \otimes D^{h}
		\, \right)
\nonumber \\
&	=	\exp \left( c f_{G} D \right) \otimes G
	+	G \otimes \exp \left( c f_{G} D \right),
\end{align}
where $c \in \mathbb{C}$ is a non-zero parameter.
The $Q$-$Q$ and $S$-$S$ twist elements and the corresponding twisted coproduct
expansion terminates at finite order in the non(anti)commutative parameter 
$C$, as discussed in reference to (\ref{coproductG}).
Terms of third order, \( \mathcal{O}(C^3) \), drop out, due to
the nilpotency of generators, such as \( Q_{\alpha} Q_{\beta} Q_{\gamma} = 0 \, 
\left( \alpha, \beta, \gamma \in \{ 1,2 \} \right) \).
Hence we can evaluate the \( S \)-\( S \) twisted coproducts as
\begin{align}
\label{EQN:SS_TwistedCoproduct}
	\Delta_{t}^{\mathrm{SS}} ( G )
&	=	\Delta_{0} ( G )
	-	\frac{1}{2} C^{\alpha\beta}
		\left( \,
			S_{\alpha} \!\otimes [ \, S_{\beta}, G \, ]_{\pm}
		+	( -1 )^{|G|} \,
			[ \, S_{\alpha}, G \, ]_{\pm} \!\otimes S_{\beta}
		\, \right)
\nonumber \\
& \qquad \qquad \,
	-	\frac{1}{8} C^{\gamma\delta} C^{\alpha\beta}
		\left( \,
			S_{\gamma} S_{\alpha} \!\otimes
			[ \, S_{\delta}, [ \, S_{\beta}, G \, ]_{\pm} \, ]_{\pm}
		+	[ \, S_{\gamma}, [ \, S_{\alpha}, G \, ]_{\pm} \, ]_{\pm}
			\!\otimes S_{\delta} S_{\beta}
		\, \right).
\end{align}
The coproducts for the generators of \( \mathcal{SC} \) are as follows:
\begin{alignat}{3}
\label{EQN:SS_TwistedCoproduct_P}
&	\Delta_{t}^{\mathrm{SS}} ( P_{\alpha\dot{\alpha}} )
& &	=	\Delta_{0}( P_{\alpha\dot{\alpha}} )
	+	\epsilon_{\alpha\beta} \, C^{\beta\gamma}
		\left( \,
			S_{\gamma} \!\otimes \bar{Q}_{\dot{\alpha}}
		+	\bar{Q}_{\dot{\alpha}} \!\otimes S_{\gamma}
		\, \right) \, ,
\\[.3em]
\label{EQN:SS_TwistedCoproduct_M}
&	\Delta_{t}^{\mathrm{SS}} ( M_{\alpha\beta} )
& &	=	\Delta_{0} ( M_{\alpha\beta} )
	+	\frac{i}{2} \, C^{\gamma\delta}
		\left( \,
			S_{\gamma} \!\otimes \epsilon_{\delta(\alpha} S_{\beta)}
		+	\epsilon_{\gamma(\alpha} S_{\beta)} \!\otimes S_{\delta}
		\, \right) \, ,
\\[.5em]
\label{EQN:SS_TwistedCoproduct_Mbar}
&	\Delta_{t}^{\mathrm{SS}} ( \bar{M}_{\dot{\alpha}\dot{\beta}} )
& &	=	\Delta_{0} ( \bar{M}_{\dot{\alpha}\dot{\beta}} ) \, ,
\\[.7em]
\label{EQN:SS_TwistedCoproduct_Q}
&	\Delta_{t}^{\mathrm{SS}} ( Q_{\alpha} )
& &	=	\Delta_{0} ( Q_{\alpha} )
	-	2i \, C^{\beta\gamma}
		\left( \,
			S_{\beta} \otimes M_{\alpha\gamma}
		-	M_{\alpha\gamma} \otimes S_{\beta}
		\, \right)
\nonumber \\[.5em]
& & & \quad
	+	i \epsilon_{\alpha\beta} \, C^{\beta\gamma}\vspace{3em}
		\left( \,
			S_{\gamma} \otimes D
		-	D \otimes S_{\gamma}
		\, \right)
	+	3 \, \epsilon_{\alpha\beta} \, C^{\beta\gamma}
		\left( \,
			S_{\gamma} \otimes R
		-	R \otimes S_{\gamma}
		\, \right)
\nonumber \\
& & & \quad
	-	\frac{1}{4}
		C^{\delta\eta} C^{\beta\gamma}
		\left[ \,
			S_{\delta} S_{\beta}
			\otimes
			\left(
				\epsilon_{\eta\gamma} S_{\alpha}
			-	2 \epsilon_{\alpha(\eta} S_{\gamma)}
			\right)
		+	\left(
				\epsilon_{\eta\gamma} S_{\alpha}
			-	2 \epsilon_{\alpha(\eta} S_{\gamma)}
			\right)
			\otimes
			S_{\delta} S_{\beta}
		\, \right] \, ,
\\[.3em]
\label{EQN:SS_TwistedCoproduct_Qbar}
&	\Delta_{t}^{\mathrm{SS}} ( \bar{Q}_{\dot{\alpha}} )
& &	=	\Delta_{0} ( \bar{Q}_{\dot{\alpha}} ) \, ,
\\[.3em]
\intertext{}
\label{EQN:SS_TwistedCoproduct_D}
&	\Delta_{t}^{\mathrm{SS}} ( D )
& &	=	\Delta_{0} ( D )
	+	\frac{i}{2} \, C^{\alpha\beta} \, S_{\alpha} \! \otimes S_{\beta} \, ,
\\
\label{EQN:SS_TwistedCoproduct_R}
&	\Delta_{t}^{\mathrm{SS}} ( R )
& &	=	\Delta_{0} ( R )
	+	\frac{1}{2} \, C^{\alpha\beta} \, S_{\alpha} \! \otimes S_{\beta} \, ,
\\[.5em]
\label{EQN:SS_TwistedCoproduct_S}
&	\Delta_{t}^{\mathrm{SS}} ( S_{\alpha} )
& &	=	\Delta_{0}( S_{\alpha} ) \, ,
\\[.7em]
\label{EQN:SS_TwistedCoproduct_Sbar}
&	\Delta_{t}^{\mathrm{SS}} ( \bar{S}_{\dot{\alpha}} )
& &	=	\Delta_{0}( \bar{S}_{\dot{\alpha}} )
	+	C^{\alpha\beta}
		\left( \,
			K_{\alpha\dot{\alpha}} \otimes S_{\beta}
		-	S_{\beta} \otimes K_{\alpha\dot{\alpha}}
		\, \right) \, ,
\\[.7em]
\label{EQN:SS_TwistedCoproduct_K}
&	\Delta_{t}^{\mathrm{SS}} ( K_{\alpha\dot{\alpha}} )
& &	=	\Delta_{0}( K_{\alpha\dot{\alpha}} ) \, .
\\[-1em] \nonumber 
\end{alignat}
A bosonic-fermionic type twist element is constructed from two kinds of generators
which span some abelian subsector, as listed in Table \ref{TBL:PossibleCombinations}.
For example, we have \( P \)-\( Q \), \( S \)-\( \bar{Q} \), \( K \)-\( S \), and so on.
Their coproducts are slightly different from the previous two.
As a typical case, we study the \( P \)-\( Q \) twist.
Following the definition given in \cite{KS},
the twist element with a Grassmann number parameter \( \lambda \) is
\begin{equation}
\label{EQN:PQ_TwistElement}
	\mathcal{F}_{\mathrm{PQ}}
	=	\exp
		\left( \,
			\frac{i}{2} \, \lambda^{m\alpha}
			\left[ \,
				P_{m} \!\otimes Q_{\alpha}
			-	Q_{\alpha} \!\otimes P_{m}
			\, \right]
		\, \right).
\end{equation}
The \( h \)-th order terms in the twisted coproduct \( \Delta_{t}^{\mathrm{PQ}} ( G ) \)
for any generators \( G \) of \( \mathcal{SC} \) are written in the form (with the notation 
\( \mathrm{ad}^{\pm}_{A} ( B ) := \left[ \, A, B \, \right]_{\pm} \))
\begin{align}
&	P_{m_{1}} \!\cdots P_{m_{h}}
	\!\otimes
	\mathrm{ad}^{\pm}_{Q_{\alpha_{1}}} \!\cdots\, \mathrm{ad}^{\pm}_{Q_{\alpha_{h}}}
	\! ( G )
\ , \quad
	Q_{\alpha_{1}} \!\cdots Q_{\alpha_{h}}
	\!\otimes
	\mathrm{ad}^{\pm}_{P_{m_{1}}} \!\cdots\, \mathrm{ad}^{\pm}_{P_{m_{h}}}
	\! ( G ),
\nonumber \\
&	P_{m_{1}} \!\cdots P_{m_{k}} Q_{\alpha_{k+1}} \!\cdots Q_{\alpha_{h}}
	\!\otimes
	\mathrm{ad}^{\pm}_{Q_{\alpha_{1}}} \!\cdots\, \mathrm{ad}^{\pm}_{Q_{\alpha_{k}}}
	\mathrm{ad}^{\pm}_{P_{m_{k+1}}} \!\cdots\, \mathrm{ad}^{\pm}_{P_{m_{h}}}
	\! ( G )
	\quad
	\forall \, k \in \{ \, 1,\cdots,h-1 \, \},
\end{align}
and the left-right reversed counterparts.
Then,for the reason mentioned above, 
all \( \mathcal{O}( \lambda^4 ) \) terms vanish in the evaluation of the 
twisted coproduct.
\par
%
\section{Summary and Discussion}
\label{SEC:summary}
In this paper we have discussed the broken and unbroken superconformal symmetries on various 
non(anti)commutative spaces in the context of the Drinfel'd twisted Hopf algebra.
The non(anti)commutative (super)spaces are realized as the representation space 
of the twisted Hopf algebra.
The non(anti)commutativity of the superspace depends on the choice of the twist element,
e.g. the $P$-$P$ twist element $\mathcal{F}_{\mathrm{PP}}$ for the canonical 
noncommutative plane $[x^m, x^n] = i \theta^{mn}$
and the \( S \)-\( S \) twist element for the non-canonical non(anti)commutative superspace
(\ref{EQN:SS_TwistedCommutator}).
A theory defined on 
a noncommutative space generally exhibits a violation of some part 
of the original (commutative space) symmetries and some part, including 
linear combinations of the generators in the broken symmetry, remains intact.
These broken and unbroken symmetries can be systematically classified 
with respect to the Drinfel'd twisted Hopf algebra.
\par
Strictly speaking, these ``broken'' symmetries are not broken in a 
certain sense, because the algebra can be recovered, provided that the corresponding coproduct
is modified, as we have seen in \S \ref{SEC:twisted_coproduct}.
\par
We also considered all possible abelian twist elements in the 
superconformal algebra and classified the structure of broken and unbroken 
symmetries corresponding to each choice of the twist element.
In addition, we showed that the $S$-$S$ twist element $\mathcal{F}_{\mathrm{SS}}$ provides
a non-canonical non(anti)commutative structure in the superspace that differs
from that of the canonical type, resulting from the $P$-$P$ or $Q$-$Q$ twist.
Nevertheless, the $S$-$S$ twisted star product can be computed explicitly,
and the effects of the deformation are of finite order in the noncommutativity parameter \( C \).
\par
In physical applications,
one might expect that the Hopf algebra approach, which induces 
non-canonical non(anti)commutativity on the (super)space
provides a useful tool to classify the symmetries of low energy effective field theories 
in superstring theory with non-constant background fields,
i.e. field theories specified by commutation relations of the type 
\begin{equation}
	[ x^{m}, x^{n} ]
	=	C^{mn} ( x, \theta, \bar{\theta} )
\ , \quad
	\{ \theta^{\alpha}, \theta^{\beta} \}
	=	C^{\alpha\beta} ( x, \theta, \bar{\theta} ),
\end{equation}
which, in general,
contain the non(anti)commutative structures presented in Table \ref{TBL:Broken_And_Unbroken}.
\par
The study carried out in this paper may provide a general method for the 
investigation of the symmetry structures of superconformal theories on noncommutative geometries.
%
%
\section*{Acknowledgements}
\label{SEC:Acknowledge}
Two of the authors Y.~K and S.~S would like to thank the member of the Helsinki 
Institute of Physics (HIP) for their hospitality while a part of this paper 
was being prepared. 
S.~S is supported by the Interactive Research Center of Science at the Tokyo Institute of Technology.

\end{document}